\documentclass[conference,a4paper]{IEEEtran}
	\newtheorem{theorem}{Theorem}
	\newtheorem{lemma}[theorem]{Lemma}
	\newtheorem{corollary}[theorem]{Corollary}
	\newtheorem{definition}[theorem]{Definition}
	\newtheorem{proposition}[theorem]{Proposition}
	\newtheorem{ivp}[theorem]{Initial Value Problem}
	
\usepackage[cmex10]{amsmath}
\usepackage{mathtools, amssymb, mleftright, booktabs}
	\def\>{\succcurlyeq}
	\def\<{\preccurlyeq}

\usepackage{tikz-cd, pgfplots}
	\pgfmathsetseed{792262833}
	
	\tikzset{every picture/.style={cap=round, join=round}}
	\pgfplotsset{compat/show suggested version=false,compat=1.17}

\usepackage[noadjust]{cite}
\usepackage{hyperref}
	\hypersetup{
		colorlinks, allcolors=green!20!blue!80!black,
		pdfsubject={Information Theory (cs.IT)},
		pdfkeywords={
			polar code,
			partial order,
			dual channel,
			ODE,
			IVP,
			Green's theorem, 
		},
	}

\begin{document}

\advance\baselineskip0pt plus0.2pt minus0.1pt

\title{Optimal Self-Dual Inequalities \\ to Order Polarized BECs}

\author{%
	\IEEEauthorblockN{Ting-Chun Lin}%
	\IEEEauthorblockA{%
		University of California San Diego, CA, USA\\
		Hai (Foxconn) Research Institute, Taipei, Taiwan\\
		til022@ucsd.edu%
	}
	\and
	\IEEEauthorblockN{Hsin-Po Wang}%
	\IEEEauthorblockA{%
		University of California, Berkeley, CA, USA\\
		simple@berkeley.edu%
	}
}
\maketitle

\begin{abstract}\boldmath\def\>{\succcurlyeq}
	$1 - (1-x^M) ^ {2^M} > (1 - (1-x)^M) ^{2^M}$ is proved for all $x \in [0,1]$
	and all $M > 1$.  This confirms a conjecture about polar code, made by Wu
	and Siegel in 2019, that $W^{0^m 1^M}$ is more reliable than $W^{1^m 0^M}$,
	where $W$ is any binary erasure channel and $M = 2^m$.  The proof relies on
	a remarkable relaxation that $m$ needs not be an integer, a cleverly crafted
	hexavariate ordinary differential equation, and a genius generalization of
	Green's theorem that concerns function composition.  The resulting
	inequality is optimal, $M$ cannot be $2^m - 1$, witnessing how far polar
	code deviates from Reed--Muller code.
\end{abstract}

\section{Introduction}

	If there is anything that contributes to polar code's outstandingly low
	decoding complexity, it is the insistence of the successive cancellation
	decoder that some information bits are processed earlier and some later
	\cite{Ari09}.  This, generally speaking, makes bits that are processed later
	better protected.  For instance, if $abc1xyz$ is the binary expansion of a
	bit's index, it has been shown that it is better protected than the bit
	whose index's binary expansion reads $abc0xyz$.  With a more elaborated
	argument, it was shown that $abc10xyz$ is better protected than $abc01xyz$
	\cite{BDO16, Sch16}.

	It is very difficult to prove otherwise, that there exists an information
	bit that is processed earlier yet better protected.  We do not know, for
	instance, whether the bit indexed by $abc011xyz$ is \emph{always} better
	protected than the bit indexed by $abc100xyz$, despite of countless
	simulations that support so (to an extent that some works advocate
	postponing $011$ and processing $100$ first \cite{BFS17, TrT21, LYH22}).  In
	this work, we give several examples over binary erasure channels (BEC) where
	a bit whose binary index is lexicographically earlier is always better
	protected.  Some examples are $011$ versus $100$, $00111$ versus $10000$,
	$01011$ versus $10100$, $001111$ versus $110000$, $000111$ versus $100000$,
	and the list goes on.

	For any BEC $W$ with capacity $I(W) = x$, let $W^0$ be the BEC with capacity
	$I(W^0) = x^2$; let $W^1$ be the BEC with capacity $I(W^1) = 1 - (1 - x)^2$.
	For any binary string $b_1 b_2 \dotsc b_n \in \{0, 1\}^n$, define $W^{b_1
	b_2 \dotsm b_n}$ as $\bigl( \dotsm (W^{b_1}) ^{b_2} \dotsm \bigr) ^{b_n}$.
	For any binary strings $a \in \{0, 1\}^m$ and $b \in \{0, 1\}^n$, we say
	that the former outperforms the latter if, for any underlying BEC $W$, the
	capacity of $W^a$ is higher than or equal to the capacity of $W^b$.  We
	denote this as $a \> b$.  When there are consecutive bits in a binary
	string, for instance $000$ and , we abbreviate it as $0^3$ and $1^5$.
	
	The main results of this work follows.  We show that, to prove a comparison
	of the form $0^m 1^n \> 1^m 0^n$, it suffices to check if the comparison
	holds for $x = 1/2$.

	\begin{theorem}[main theorem]                               \label{thm:main}
		Let $m$ and $n$ be positive numbers.  We have $0^m 1^n \> 1^m 0^n$ if
		and only if we have  $(1 - 1/2^{2^m}) ^{2^n} \leq 1/2$.
	\end{theorem}

	Theorem~\ref{thm:main} resolves a conjecture by Wu and Siegel, which states
	that $0^m 1^{2^m} \> 1^m 0^{2^m}$ \cite[(45)]{WuS19}, as $(1 - 1/2^{2^m}) ^
	{2^{2^m}}$ is indeed $\leq 1/2$.  In fact, Theorem~\ref{thm:main} implies
	something stronger: that $0^m 1^{2^m-0.528} \> 1^m 0^{2^m-0.528}$ for all $m
	> 0$.  We hope that readers are not turned away by the unexpected notion
	that there are $2^m-0.528$ ones on the left-hand side and $2^m-0.528$ zeros
	on the right-hand side.  We never said $m, n$ are integers.  Making sense of
	non-integer amounts of zeros and ones is the breakthrough that enables the
	proof of the main theorem.

	This paper is organized as follows.  Section~\ref{sec:continuous} defines
	what $0^p$ and $1^q$ mean for real numbers $p$ and $q$.
	Section~\ref{sec:10>01} demonstrates the key idea of the main theorem using
	an example---we will prove that $1^p 0^q \> 0^p 1^q$ for any real numbers
	$p, q > 0$.  Section~\ref{sec:local} discusses $0^p 1^q \> 1^r 0^s$ for
	small $p, q, r, s$.  Section~\ref{sec:global} generalizes Green's theorem to
	discuss the case when $p, q, r, s$ are not small.  Section~\ref{sec:main}
	proves the main theorem and some consequences.

\section{Interpolating the Action of Squaring}            \label{sec:continuous}

	In this section, we parametrize the action of squaring a number so it makes
	sense to say ``I am $61.8\%$ done with squaring this number.''

	For any real numbers $p$ and $q$, define $I_0^p(x) \coloneqq x^{2^p}$ and
	$I_1^q(x) \coloneqq 1 - (1 - x)^{2^q}$.  Note that $I_0^0(x)$ is just $x$,
	that $I_0^1(x)$ is just $x^2$, and that $I_0^r (I_0^p (x) )$ is just
	$I_0^{p+r}(x)$.  Consider real numbers $p_1, p_2, \dotsc, p_n$ and $q_1,
	q_2, \dotsc, q_n$.  Let a \emph{realistic string} be of the form $0^{p_1}
	1^{q_1} 0^{p_2} 1^{q_2} \dotsm 0^{p_n} 1^{q_n}$ and correspond to the
	function composition
	\begin{equation}                                             \label{for:Ipq}
		\def\1{I_0^{p_1}(x)}
		\def\2{I_1^{q_1}(\1)}
		\def\3{I_0^{p_2} \bigl( \2 \bigr)}
		\def\4{I_1^{q_2} \bigl( \3 \bigr)}
		\def\5{\dotsm \4 \dotsm}
		\def\6{I_0^{p_n} \Bigl( \5 \Bigr)}
		\def\7{I_1^{q_n} \Bigl( \6 \Bigr)}
		\7
	\end{equation}
	We say $0^{p_1} 1^{q_1} 0^{p_2} 1^{q_2} \dotsm 0^{p_n} 1^{q_n} \> 0^{r_1}
	1^{s_1} 0^{r_2} 1^{s_2} \dotsm 0^{r_n} 1^{s_n}$ if formula~\eqref{for:Ipq}
	is greater than or equal to
	\[
		\def\1{I_0^{r_1}(x)}
		\def\2{I_1^{s_1}(\1)}
		\def\3{I_0^{r_2} \bigl( \2 \bigr)}
		\def\4{I_1^{s_2} \bigl( \3 \bigr)}
		\def\5{\dotsm \4 \dotsm}
		\def\6{I_0^{r_n} \Bigl( \5 \Bigr)}
		\def\7{I_1^{s_n} \Bigl( \6 \Bigr)}
		\7
	\]
	for all $x \in [0, 1]$.

	For all intents and purposes, we can safely ignore $0^0$ and $1^0$.  We can
	also identify $0^p 0^r$ as $0^{p+r}$ and $1^q 1^s$ as $1^{q+s}$.  Through
	these simplifications, the notion of realistic string covers all possible
	concatenations of $0^p$ and $1^q$.

	Immediately one sees that $I_0^p(x) < x < I_1^q(x)$ for all $p, q > 0$ and
	$0 < x < 1$.  Accordingly, one writes $0^p \<$ empty string $\< 1^q$.  This
	is the continuous version of $0 \<$ empty string $\< 1$.  The next section
	proves the continuous version of $01 \< 10$, illustrating our idea for the
	main theorem.

\section{\texorpdfstring{$1^p 0^q \> 0^q 1^p$}{1\^{}p 0\^{}q > 0\^{}q 1\^{}p}:
         A Demonstration of Ideas}                             \label{sec:10>01}

	In this section, we want to show that for any pairs of positive real numbers
	$(p, q)$ it holds that $1^p 0^q \> 0^q 1^p$.  This serves as a toy example
	before we attack the main theorem.

	To begin, we only care for very small $p$ and $q$.  That encourages us to
	consider the Taylor expansion of $I_0^q(I_1^p(x)) - I_1^p(I_0^q(x))$ at $(p,
	q) = (0, 0)$ while treating $x$ as a constant.  The expansion looks like
	\begin{align*}
		\kern1em&\kern-1em
		I_0^q(I_1^p(x)) - I_1^p(I_0^q(x)) \\
		& = \ln(2)^2
		    \Bigl( \ln(x)\ln(1{-}x) - x\ln(x) - (1{-}x)\ln(1{-}x) \Bigr)
		    \cdot pq \\
		&\kern1em \pm O(|p|^3 + |q|^3).
	\end{align*}
	Clearly the term between the big parentheses is positive whenever $0 < x <
	1$.  Thus, for any $x \in [0, 1]$, the difference $I_0^q(I_1^p(x)) -
	I_1^p(I_0^q(x))$ is positive provided that $p$ and $q$ are positive yet
	small enough.

	The punchline is that $p$ and $q$ need not be small.

	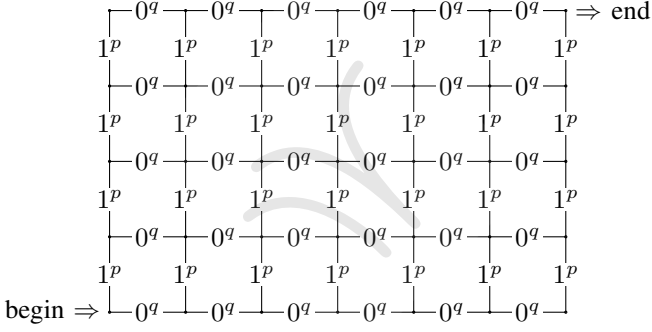
\begin{figure}
		\centering
		\begin{tikzpicture}
			\def\H{4}\pgfmathtruncatemacro\h{\H-1}
			\def\W{6}\pgfmathtruncatemacro\w{\W-1}
			\draw (0,0) node[left]{begin $\Rightarrow$};
			\draw (\W,\H) node[right]{$\Rightarrow$ end};
			\foreach\x in{0,...,\W}{
				\foreach\y in{0,...,\H}{
					\draw (\x,\y)circle(.4pt);
				}
			}
			\tikzset{every node/.style={fill=white,inner sep=1pt}}
			\foreach\x in{0,...,\w}{
				\foreach\y in{0,...,\H}{
					\draw (\x,\y) --node{$0^q$} +(1,0);
				}
			}
			\foreach\x in{0,...,\W}{
				\foreach\y in{0,...,\h}{
					\draw (\x,\y) --node{$1^p$} +(0,1);
				}
			}
			\path[overlay] (\W/2,\H/2)
				node[rotate=-45,opacity=.1,scale=10]{$\,\>$};
		\end{tikzpicture}
		\caption{Green's theorem and inequality: Suppose that every small square
		is such that going up-and-then-right results in a better reward than
		going right-and-then-up.  Then the path that travels from the southwest
		corner to the northeast corner via the northwest corner yields better
		reward than that via the southeast corner.  That is to says, $1^{4p}
		0^{6q} \> 0^{6q} 1^{4p}$.}                             \label{fig:10>01}
	\end{figure}

	Inspect Figure~\ref{fig:10>01}.  In this figure, we want to travel from the
	southwest corner to the northeast corner.  Each possible path corresponds to
	a realistic string.  For instance, two rights, one up, four rights, and
	three ups means  $0^{2q} 1^p 0^{4q} 1^{3p}$.  The inequality $1^p 0^q \> 0^q
	1^p$ translates into a gimmick that, if a path lies northwest to another
	path, for instance $1^p 0^{2q} 1^{3p} 0^{4q}$ is northwest to $0^{4q} 1^{3p}
	0^{2q} 1^p$, then the former $\>$ the latter, for instance $1^p 0^{2q}
	1^{3p} 0^{4q} \> 0^{4q} 1^{3p} 0^{2q} 1^p$.  Since Figure~\ref{fig:10>01}
	can be made arbitrarily large with arbitrarily small squares, we conclude
	the following.

	\begin{theorem}[continuous version of $10 \> 01$]
		For any real numbers $p$ and $q$ that are positive, $1^p 0^q \> 0^q
		1^p$.  That is, $I_0^q(I_1^p(x)) \geq I_1^p(I_0^q(x)) $ for all $0 \leq
		x \leq 1$.
	\end{theorem}

	The proof of the main theorem will follow the same logic.  We first work out
	a local necessary condition for $I_1^q(I_0^p(x)) - I_0^s(I_1^r(x)) > 0$, and
	then we construct a big, fine grid on which we compare paths.

\section{Local Positivity for \texorpdfstring{$0^p 1^q \> 1^r 0^s$}
		 {0\^{}p 1\^{}q > 1\^{}r 0\^{}s}}                      \label{sec:local}

	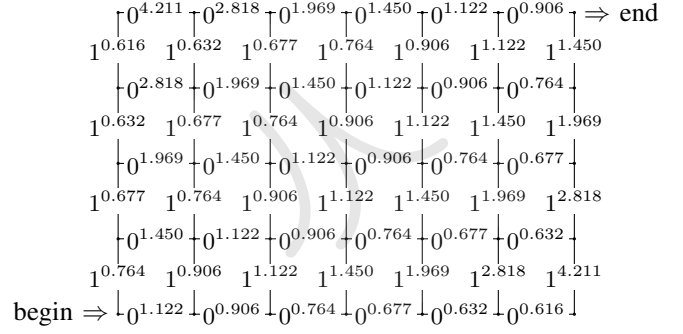
\begin{figure}
		\centering
		\begin{tikzpicture}
			\def\H{4}\pgfmathtruncatemacro\h{\H-1}
			\def\W{6}\pgfmathtruncatemacro\w{\W-1}
			\draw (0,0) node[left]{begin $\Rightarrow$};
			\draw (\W,\H) node[right]{$\Rightarrow$ end};
			\def\0{0.616}
			\def\1{0.632}
			\def\2{0.677}
			\def\3{0.764}
			\def\4{0.906}
			\def\5{1.122}
			\def\6{1.450}
			\def\7{1.969}
			\def\8{2.818}
			\def\9{4.211}
			\foreach\x in{0,...,\W}{
				\foreach\y in{0,...,\H}{
					\draw (\x,\y)circle(.4pt);
				}
			}
			\tikzset{every node/.style={fill=white,inner sep=1pt}}
			\foreach\x in{0,...,\w}{
				\foreach\y in{0,...,\H}{
					\xdef\p{\csname\the\numexpr\y-\x+5\endcsname}
					\draw (\x,\y) --node{\small$0^{\p}$} +(1,0);
				}
			}
			\foreach\x in{0,...,\W}{
				\foreach\y in{0,...,\h}{
					\xdef\q{\csname\the\numexpr\x-\y+3\endcsname}
					\draw (\x,\y) --node{\small$1^{\q}$} +(0,1);
				}
			}
			\path[overlay] (\W/2,\H/2)
				node[rotate=-45,opacity=.1,scale=10]{$\<\,$};
		\end{tikzpicture}
		\caption{Green's theorem and inequality: Similar to
		Figure~\ref{fig:10>01} except that, here, the assumption is that going
		right-and-then-up is better.  Thus the path via the southeast corner
		yields the highest reward among all paths.  While numbers in this figure
		are found by a computer, we will prove mathematically that an ideal grid
		exists.}                                               \label{fig:01>10}
	\end{figure}

	This section paves the way toward Theorem~\ref{thm:main}, the main theorem.
	More paving works will be done in the Section~\ref{sec:global}, whereas the
	proof of the main theorem will be completed in Section~\ref{sec:main}.  For
	the present section, our goal is to characterize the asymptotic behavior of
	the quadruples $(p, q, r, s)$ such that $0^p 1^q \> 1^r 0^s$.

	Applying Taylor approximation at the neighborhood of $(p, q, r, s) = (0, 0,
	0, 0)$, we obtain
	\begin{align*}
		\kern1em&\kern-1em
		I_1^q(I_0^p(x)) - I_0^s(I_1^r(x)) \\
		& =         - \ln(2) x \ln(x) \cdot (s-p) \\
		&\kern1.1em - \ln(2) (1-x) \ln(1-x) \cdot(q-r) \\
		&\kern1.1em + \ln(2)^2 x \ln(x) (1+\ln(1-x)) \cdot pq \\
		&\kern1.1em + \ln(2)^2 (1-x) \ln(1-x) (1+\ln x) \cdot rs \\
		&\kern1.1em \pm O(|p|^3 + |q|^3 + |r|^3 + |s|^3).           
	\end{align*}
	To comprehend when the difference is nonnegative, we first want to
	understand when the Taylor approximation is barely nonnegative.  That is to
	ask, what $(p, q, r, s)$ makes $\Delta(y) = 0$ for some $y \in (0, 1)$ and
	$\Delta(x) > 0$ for any other $x \in (0, 1) \setminus y$, where
	\begin{align*}
		\Delta(x)
		& \coloneqq - \ln(2) x \ln(x) \cdot (s-p) \\
		&\kern1.3em - \ln(2) (1-x) \ln(1-x) \cdot(q-r) \\
		&\kern1.3em + \ln(2)^2 x \ln(x) (1+\ln(1-x)) \cdot pq \\
		&\kern1.3em + \ln(2)^2 (1-x) \ln(1-x) (1+\ln x) \cdot rs?  
	\end{align*}

	To proceed further, let us limit ourselves to the case where $s - p$ and $q
	- r$ are both of quadratic order.  There is no rigorous justification why
	this should be the case.  It just $feels$ $right$, and it nevertheless leads
	to a rigorous proof of the main theorem.  To elaborate, $s - p$ should be a
	polynomial in $p, q, r, s$ without constant and linear term; similarly, $q -
	r$ should be a polynomial in $p, q, r, s$ without constant and linear term;
	Henceforth,
	\[
		rs
		= \bigl( p + O(\text{quadratic}) \bigr)
		  \bigl( q - O(\text{quadratic}) \bigr)
		= pq + O(\text{cubic}).
	\]
	We further limit ourselves to the case that $s - p$, $q - r$, $pq$, and $rs$
	are multiples of each other up to some cubic error.  Again, we cannot
	explain why it must be like that; just bear with us.

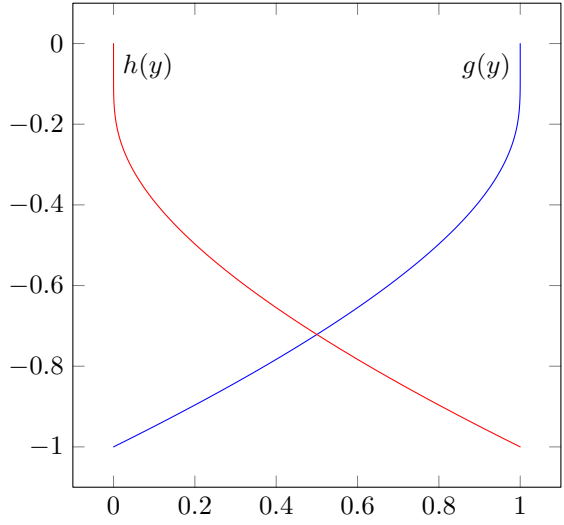
\begin{figure}
		\centering
		\begin{tikzpicture}
			\begin{axis}[height=8cm, width=8cm, no marks]
				\addplot [blue] table [x=g, y=-y] {\GHJK};
				\node at (1, 0) [below left] {$g(y)$};
				\addplot [red] table [x=h, y=-y] {\GHJK};
				\node at (0, 0) [below right] {$h(y)$};
			\end{axis}
		\end{tikzpicture}
		\caption{Two of the coefficients in equation~\eqref{equ:linear}.
		}                                                    \label{fig:g and h}
	\end{figure}

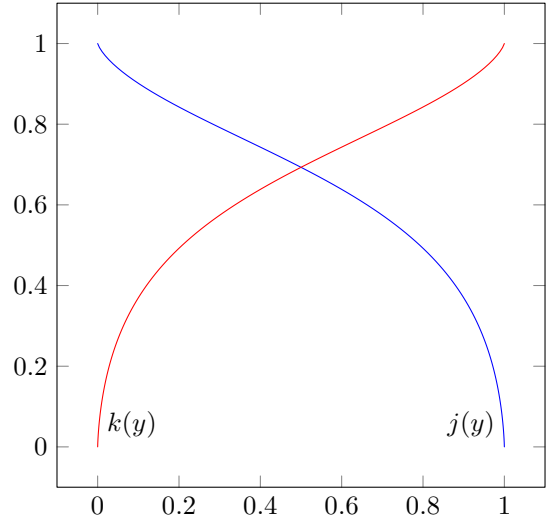
\begin{figure}
		\centering
		\begin{tikzpicture}
			\begin{axis}[height=8cm, width=8cm, no marks]
				\addplot [blue] table [x=j, y=+y] {\GHJK};
				\node at (1, 0) [above left, align=right] {$j(y)$};
				\addplot [red] table [x=k, y=+y] {\GHJK};
				\node at (0, 0) [above right, align=left] {$k(y)$};
			\end{axis}
		\end{tikzpicture}
		\caption{The solution of equation~\eqref{equ:linear} as functions in
		$y$.}                                                \label{fig:j and k}
	\end{figure}

	Let $j$ be $1 - (s - p) / \ln(2) pq$.  Let $k$ be $1 - (q - r) / \ln(2) pq$.
	This means that we can replace $s - p$ by $\ln(2) (1 - j) pq$ and $q - r$ by
	$\ln(2) (1 - k) pq$.  The delta we want to make zero then simplifies to
	\begin{align*}
		\Delta(x)
		&= \ln(2)^2 \ln(x) \ln(1-x) \cdot pq \\
		&\kern0.1em  + \ln(2)^2 x \ln(x) j \cdot pq \\
		&\kern0.1em  + \ln(2)^2 (1-x) \ln(1-x) k \cdot pq.
	\end{align*}
	Factoring out $\ln(2)^2 \ln(x) \ln(1-x) pq$, we conclude that the quadratic
	approximation is zero at $x = y$ if
	\[ 1 + \frac{jy}{\ln(1-y)} + \frac{k(1-y)}{\ln y} = 0. \]
	See Figure~\ref{fig:g and h} for the plots of coefficients of $j$ and $k$.
	See Figure~\ref{fig:j and k} for the plots of $j$ and $k$

	\pgfplotstableread{
		g        h       -y        j        k        +y
		1.       0.       0.       1.       0.       0.      
		1.       0.      -0.01     0.9997   0.0003   0.01    
		1.       0.      -0.02     0.99922  0.00078  0.02    
		1.       0.      -0.03     0.99863  0.00137  0.03    
		1.       0.      -0.04     0.99793  0.00207  0.04    
		1.       0.      -0.05     0.99712  0.00288  0.05    
		1.       0.      -0.06     0.99621  0.00379  0.06    
		1.       0.      -0.07     0.99519  0.00481  0.07    
		1.       0.      -0.08     0.99407  0.00593  0.08    
		0.99999  0.00001 -0.09     0.99285  0.00715  0.09    
		0.99995  0.00005 -0.1      0.99152  0.00848  0.1     
		0.99989  0.00011 -0.11     0.99009  0.00991  0.11    
		0.99976  0.00024 -0.12     0.98854  0.01146  0.12    
		0.99954  0.00046 -0.13     0.98687  0.01313  0.13    
		0.99921  0.00079 -0.14     0.98509  0.01491  0.14    
		0.99872  0.00128 -0.15     0.98319  0.01681  0.15    
		0.99805  0.00195 -0.16     0.98116  0.01884  0.16    
		0.99716  0.00284 -0.17     0.979    0.021    0.17    
		0.99605  0.00395 -0.18     0.9767   0.0233   0.18    
		0.99467  0.00533 -0.19     0.97427  0.02573  0.19    
		0.99302  0.00698 -0.2      0.97169  0.02831  0.2     
		0.99108  0.00892 -0.21     0.96897  0.03103  0.21    
		0.98883  0.01117 -0.22     0.96608  0.03392  0.22    
		0.98627  0.01373 -0.23     0.96304  0.03696  0.23    
		0.98338  0.01662 -0.24     0.95983  0.04017  0.24    
		0.98017  0.01983 -0.25     0.95645  0.04355  0.25    
		0.97663  0.02337 -0.26     0.95289  0.04711  0.26    
		0.97275  0.02725 -0.27     0.94915  0.05085  0.27    
		0.96854  0.03146 -0.28     0.94521  0.05479  0.28    
		0.964    0.036   -0.29     0.94107  0.05893  0.29    
		0.95912  0.04088 -0.3      0.93672  0.06328  0.3     
		0.95391  0.04609 -0.31     0.93216  0.06784  0.31    
		0.94837  0.05163 -0.32     0.92737  0.07263  0.32    
		0.94251  0.05749 -0.33     0.92236  0.07764  0.33    
		0.93632  0.06368 -0.34     0.9171   0.0829   0.34    
		0.92981  0.07019 -0.35     0.91159  0.08841  0.35    
		0.92299  0.07701 -0.36     0.90582  0.09418  0.36    
		0.91586  0.08414 -0.37     0.89978  0.10022  0.37    
		0.90842  0.09158 -0.38     0.89347  0.10653  0.38    
		0.90068  0.09932 -0.39     0.88686  0.11314  0.39    
		0.89264  0.10736 -0.4      0.87996  0.12004  0.4     
		0.88431  0.11569 -0.41     0.87274  0.12726  0.41    
		0.87569  0.12431 -0.42     0.86521  0.13479  0.42    
		0.86678  0.13322 -0.43     0.85734  0.14266  0.43    
		0.85759  0.14241 -0.44     0.84913  0.15087  0.44    
		0.84813  0.15187 -0.45     0.84056  0.15944  0.45    
		0.83839  0.16161 -0.46     0.83163  0.16837  0.46    
		0.82839  0.17161 -0.47     0.82232  0.17768  0.47    
		0.81812  0.18188 -0.48     0.81263  0.18737  0.48    
		0.8076   0.1924  -0.49     0.80253  0.19747  0.49    
		0.79681  0.20319 -0.5      0.79202  0.20798  0.5     
		0.78578  0.21422 -0.51     0.78109  0.21891  0.51    
		0.7745   0.2255  -0.52     0.76973  0.23027  0.52    
		0.76297  0.23703 -0.53     0.75793  0.24207  0.53    
		0.7512   0.2488  -0.54     0.74568  0.25432  0.54    
		0.7392   0.2608  -0.55     0.73297  0.26703  0.55    
		0.72696  0.27304 -0.56     0.7198   0.2802   0.56    
		0.7145   0.2855  -0.57     0.70615  0.29385  0.57    
		0.70181  0.29819 -0.58     0.69203  0.30797  0.58    
		0.68889  0.31111 -0.59     0.67743  0.32257  0.59    
		0.67576  0.32424 -0.6      0.66235  0.33765  0.6     
		0.66241  0.33759 -0.61     0.64679  0.35321  0.61    
		0.64884  0.35116 -0.62     0.63076  0.36924  0.62    
		0.63507  0.36493 -0.63     0.61426  0.38574  0.63    
		0.62109  0.37891 -0.64     0.59729  0.40271  0.64    
		0.6069   0.3931  -0.65     0.57988  0.42012  0.65    
		0.59251  0.40749 -0.66     0.56202  0.43798  0.66    
		0.57793  0.42207 -0.67     0.54376  0.45624  0.67    
		0.56315  0.43685 -0.68     0.52509  0.47491  0.68    
		0.54817  0.45183 -0.69     0.50606  0.49394  0.69    
		0.533    0.467   -0.7      0.48669  0.51331  0.7     
		0.51765  0.48235 -0.71     0.46701  0.53299  0.71    
		0.50211  0.49789 -0.72     0.44706  0.55294  0.72    
		0.48638  0.51362 -0.73     0.42689  0.57311  0.73    
		0.47048  0.52952 -0.74     0.40653  0.59347  0.74    
		0.45439  0.54561 -0.75     0.38603  0.61397  0.75    
		0.43813  0.56187 -0.76     0.36546  0.63454  0.76    
		0.4217   0.5783  -0.77     0.34486  0.65514  0.77    
		0.40509  0.59491 -0.78     0.32429  0.67571  0.78    
		0.38832  0.61168 -0.79     0.3038   0.6962   0.79    
		0.37137  0.62863 -0.8      0.28347  0.71653  0.8     
		0.35426  0.64574 -0.81     0.26336  0.73664  0.81    
		0.33698  0.66302 -0.82     0.24352  0.75648  0.82    
		0.31955  0.68045 -0.83     0.22402  0.77598  0.83    
		0.30195  0.69805 -0.84     0.20492  0.79508  0.84    
		0.28419  0.71581 -0.85     0.18628  0.81372  0.85    
		0.26628  0.73372 -0.86     0.16817  0.83183  0.86    
		0.24821  0.75179 -0.87     0.15064  0.84936  0.87    
		0.22999  0.77001 -0.88     0.13374  0.86626  0.88    
		0.21162  0.78838 -0.89     0.11754  0.88246  0.89    
		0.1931   0.8069  -0.9      0.10207  0.89793  0.9     
		0.17443  0.82557 -0.91     0.08739  0.91261  0.91    
		0.15562  0.84438 -0.92     0.07355  0.92645  0.92    
		0.13665  0.86335 -0.93     0.06059  0.93941  0.93    
		0.11755  0.88245 -0.94     0.04856  0.95144  0.94    
		0.0983   0.9017  -0.95     0.0375   0.9625   0.95    
		0.07892  0.92108 -0.96     0.02747  0.97253  0.96    
		0.05939  0.94061 -0.97     0.01854  0.98146  0.97    
		0.03973  0.96027 -0.98     0.0108   0.9892   0.98    
		0.01993  0.98007 -0.99     0.00443  0.99557  0.99    
		0.       1.      -1.       0.       1.       1.       
	}\GHJK

	Solving $\Delta(y) = 0$ for $y$ is just the first half toward the
	characterization of $(p, q, r, s)$ such that $\Delta(x) \geq 0$.  The other
	equally important half is $\Delta'(y) = 0$.  Together, we want to solve
	\begin{equation}                                          \label{equ:linear}
		\begin{bmatrix}
			g  & h  \\
			g' & h'   
		\end{bmatrix}
		\begin{bmatrix}
			j \\ k
		\end{bmatrix}
		=
		\begin{bmatrix}
			-1 \\ 0
		\end{bmatrix},
		\kern1em\text{where}\kern1em
		\begin{cases}
			g(y) \coloneqq \frac y{\ln(1-y)}, \\
			h(y) \coloneqq \frac{1-y}{\ln y}.
		\end{cases}
	\end{equation}
	Through this linear equation we encode the so-called \emph{first-derivative
	test}.  But the real question is whether $\Delta''(y) > 0$ or else it could
	be a local maximum.  The answer is positive and is implied by the following
	four lemmas.

	\begin{lemma}[$g''$ and $h''$]                       \label{lem:g'' and h''}
		$g(y)$ and $h(y)$ are strictly convex in $y$.
	\end{lemma}

	\begin{lemma}[$g'$ and $h'$]                           \label{lem:g' and h'}
		$g(y)$ is strictly monotonically increasing; $h(y)$ is strictly
		monotonically decreasing.
	\end{lemma}

	These two are clear from Figure~\ref{fig:g and h}, rigorous proofs found in
	appendices \ref{pf:g'' and h''} and \ref{pf:g' and h'}.

	Over the unit interval $(0, 1)$, the functions $g$ and $h$ form an
	\emph{extended Chebyshev system} for the following reason: If we consider a
	linear combination $ag(y) + bh(y)$ where $ab > 0$, then the convexities of
	$g(y)$ and $h(y)$ imply that $ag(y) + bh(y)$ has at most two roots.  If we
	consider $ag(y) + bh(y)$ where $ab < 0$, then it is either increasing or
	decreasing, guaranteeing a single root.  In either case, $ag(y) + bh(y)$ has
	at most two roots.  Hence the Wronskian (the determinant of the matrix in
	\eqref{equ:linear}) never collapses to zero.

	We hereby conclude that for any $y \in (0, 1)$ there exist numbers $j$ and
	$k$ such that equation~\eqref{equ:linear} is met.  We view the solution $(j,
	k)$ as functions in $y$.

	\begin{lemma}[$j$ and $k$]                               \label{lem:j and k}
		$j(y)$ and $k(y)$ are positive whenever $0 < y < 1$.
	\end{lemma}

	\begin{lemma}[$j'$ and $k'$]                           \label{lem:j' and k'}
		$j(y)$ strictly monotonically decreases from $1$ to $0$ as $y$ goes from
		$0$ to $1$.  Meanwhile, $k(y)$ strictly monotonically increases from $0$
		to $1$ as $y$ goes from $0$ to $1$.
	\end{lemma}

	These properties are clear from Figure~\ref{fig:j and k}, rigorous proofs
	found in appendices \ref{pf:j and k} and \ref{pf:j' and k'}.

	\begin{proposition}
		For any $x, y \in (0, 1)$,
		\[ 1 + \frac{j(y)x}{\ln(1-x)} + \frac{k(y)(1-x)}{\ln x} \geq 0. \]
		The equality holds iff $x = y$.
	\end{proposition}

	\begin{IEEEproof}
		Since $j$ and $k$ are positive and $g$ and $h$ are convex, $j(y)g(x) +
		k(y)h(x)$ will be convex in $x$.  By the definitions of $j$ and $k$, we
		know $j(y)g(x) + k(y)h(x)$ has both the evaluation and the first
		derivative zero at $x = y$.  Therefore at $x = y$ is a local minimum and
		for everywhere else, i.e., $x \neq y $, the evaluation is positive.
	\end{IEEEproof}

	\begin{proposition}
		Fix any $x, y \in (0, 1)$.  Let $p, q, r, s$ be such that $s - p =
		\ln(2) (1 - j(y)) pq$ and $q - r = \ln(2) (1 - k(y)) pq$.  Then
		$\Delta(x) \geq 0$.  The equality holds iff $x = y$.
	\end{proposition}

	\begin{IEEEproof}
		Put the factor $\ln(2)^2 \ln(x) \ln(1-x) pq$ back.
	\end{IEEEproof}

	Here is a recap of what this section has so far.  We inspect the Taylor
	expansion of $I_1^q(I_0^p(x)) - I_0^s(I_1^r(x))$ and extract a condition,
	equation~\eqref{equ:linear}, that we believe will lead to the positivity of
	$\Delta$.  Via analyzing the derivatives we do see that $\Delta(y) = 0$
	granted that \eqref{equ:linear} is met; and if we plug-in any $x$ other than
	$y$ then $\Delta(x) > 0$.  It is now time to apply these knowledge to
	construct a grid.

\section{Green's Theorem for \texorpdfstring{$0^p 1^q \> 1^p 0^q$}
		 {0\^{}p 1\^{}q > 1\^{}p 0\^{}q}}                     \label{sec:global}

	\pgfplotstableread{
		t        Y        Z        P        Q        J        K
		0.       0.5      0.5      1.       1.       0.69315  0.69315
		0.1      0.47599  0.52401  0.97916  1.02171  0.70545  0.68058
		0.2      0.45206  0.54794  0.95914  1.04431  0.7175   0.66773
		0.3      0.4283   0.5717   0.93992  1.06786  0.7293   0.65461
		0.4      0.40477  0.59523  0.92146  1.09239  0.74086  0.6412 
		0.5      0.38156  0.61844  0.90375  1.11794  0.75218  0.62749
		0.6      0.35873  0.64127  0.88675  1.14458  0.76327  0.61348
		0.7      0.33637  0.66363  0.87043  1.17235  0.77412  0.59916
		0.8      0.31453  0.68547  0.85479  1.2013   0.78476  0.58454
		0.9      0.29328  0.70672  0.83979  1.23151  0.79517  0.56959
		1.       0.27267  0.72733  0.82541  1.26303  0.80536  0.55433
		1.1      0.25275  0.74725  0.81165  1.29594  0.81533  0.53874
		1.2      0.23357  0.76643  0.79848  1.33032  0.82508  0.52284
		1.3      0.21516  0.78484  0.78589  1.36624  0.83461  0.50662
		1.4      0.19756  0.80244  0.77386  1.40379  0.84392  0.49009
		1.5      0.1808   0.8192   0.76238  1.44306  0.853    0.47326
		1.6      0.16489  0.83511  0.75143  1.48416  0.86185  0.45615
		1.7      0.14984  0.85016  0.74101  1.52718  0.87046  0.43877
		1.8      0.13567  0.86433  0.73111  1.57224  0.87884  0.42113
		1.9      0.12237  0.87763  0.7217   1.61946  0.88697  0.40328
		2.       0.10995  0.89005  0.71279  1.66897  0.89484  0.38522
		2.1      0.09838  0.90162  0.70436  1.7209   0.90246  0.36701
		2.2      0.08767  0.91233  0.6964   1.77538  0.90981  0.34867
		2.3      0.07777  0.92223  0.68889  1.83258  0.91688  0.33025
		2.4      0.06869  0.93131  0.68185  1.89265  0.92367  0.31181
		2.5      0.06038  0.93962  0.67524  1.95576  0.93017  0.29338
		2.6      0.05281  0.94719  0.66906  2.02208  0.93637  0.27503
		2.7      0.04596  0.95404  0.6633   2.09179  0.94227  0.25683
		2.8      0.03978  0.96022  0.65795  2.16509  0.94786  0.23883
		2.9      0.03425  0.96575  0.653    2.24217  0.95313  0.22111
		3.       0.02931  0.97069  0.64843  2.32325  0.95808  0.20374
		3.1      0.02493  0.97507  0.64424  2.40854  0.96272  0.18679
		3.2      0.02108  0.97892  0.64041  2.49825  0.96702  0.17033
		3.3      0.0177   0.9823   0.63692  2.59263  0.97101  0.15444
		3.4      0.01476  0.98524  0.63377  2.6919   0.97468  0.13919
		3.5      0.01222  0.98778  0.63093  2.7963   0.97804  0.12463
		3.6      0.01004  0.98996  0.62839  2.90608  0.98109  0.11084
		3.7      0.00818  0.99182  0.62613  3.02148  0.98383  0.09786
		3.8      0.00661  0.99339  0.62414  3.14276  0.98629  0.08574
		3.9      0.0053   0.9947   0.6224   3.27018  0.98848  0.07451
		4.       0.00421  0.99579  0.62089  3.40398  0.9904   0.0642 
		4.1      0.00331  0.99669  0.61958  3.54445  0.99208  0.05481
		4.2      0.00258  0.99742  0.61847  3.69183  0.99353  0.04635
		4.3      0.00198  0.99802  0.61753  3.8464   0.99477  0.03879
		4.4      0.00151  0.99849  0.61674  4.00843  0.99581  0.03213
		4.5      0.00114  0.99886  0.61609  4.17819  0.99669  0.02631
		4.6      0.00085  0.99915  0.61555  4.35598  0.99742  0.02129
		4.7      0.00062  0.99938  0.61512  4.54209  0.99801  0.01702
		4.8      0.00045  0.99955  0.61477  4.73682  0.99848  0.01343
		4.9      0.00032  0.99968  0.6145   4.94048  0.99886  0.01045
		5.       0.00023  0.99977  0.61429  5.1534   0.99916  0.00802
		5.1      0.00016  0.99984  0.61413  5.37593  0.99939  0.00606
		5.2      0.00011  0.99989  0.61401  5.60844  0.99956  0.00451
		5.3      0.00007  0.99993  0.61392  5.8513   0.99969  0.0033 
		5.4      0.00005  0.99995  0.61385  6.10492  0.99979  0.00237
		5.5      0.00003  0.99997  0.6138   6.36973  0.99985  0.00168
		5.6      0.00002  0.99998  0.61377  6.64618  0.9999   0.00116
		5.7      0.00001  0.99999  0.61375  6.93474  0.99994  0.00079
		5.8      0.00001  0.99999  0.61373  7.23592  0.99996  0.00053
		5.9      0.       1.       0.61372  7.55024  0.99997  0.00034
		6.       0.       1.       0.61372  7.87826  0.99998  0.00022
		6.1      0.       1.       0.61371  8.22057  0.99999  0.00014
		6.2      0.       1.       0.61371  8.57777  0.99999  0.00008
		6.3      0.       1.       0.61371  8.95051  1.       0.00005
		6.4      0.       1.       0.61371  9.33945  1.       0.00003
		6.5      0.       1.       0.61371  9.7453   1.       0.00002
		6.6      0.       1.       0.61371 10.16879  1.       0.00001
		6.7      0.       1.       0.61371 10.61069  1.       0.     
		6.8      0.       1.       0.61371 11.0718   1.       0.     
		6.9      0.       1.       0.61371 11.55294  1.       0.     
		7.       0.       1.       0.61371 12.05499  1.       0.     
		7.1      0.       1.       0.61371 12.57886  1.       0.     
		7.2      0.       1.       0.61371 13.12549  1.       0.     
		7.3      0.       1.       0.61371 13.69588  1.       0.     
		7.4      0.       1.       0.61371 14.29105  1.       0.     
		7.5      0.       1.       0.61371 14.91209  1.       0.     
		7.6      0.       1.       0.61371 15.56012  1.       0.     
		7.7      0.       1.       0.61371 16.23631  1.       0.     
		7.8      0.       1.       0.61371 16.94189  1.       0.     
		7.9      0.       1.       0.61371 17.67812  1.       0.     
		8.       0.       1.       0.61371 18.44635  1.       0.     
		8.1      0.       1.       0.61371 19.24797  1.       0.     
		8.2      0.       1.       0.61371 20.08442  1.       0.     
		8.3      0.       1.       0.61371 20.95722  1.       0.     
		8.4      0.       1.       0.61371 21.86795  1.       0.     
		8.5      0.       1.       0.61371 22.81826  1.       0.     
		8.6      0.       1.       0.61371 23.80986  1.      -0.     
		8.7      0.       1.       0.61371 24.84456  1.      -0.     
		8.8      0.       1.       0.61371 25.92421  1.      -0.     
		8.9      0.       1.       0.61371 27.05079  1.      -0.     
		9.       0.       1.       0.61371 28.22633  1.      -0.     
		9.1      0.       1.       0.61371 29.45295  1.      -0.     
		9.2      0.       1.       0.61371 30.73287  1.      -0.     
		9.3      0.       1.       0.61371 32.06842  1.      -0.     
		9.4      0.       1.       0.61371 33.462    1.      -0.     
		9.5      0.       1.       0.61371 34.91614  1.      -0.     
		9.6      0.       1.       0.61371 36.43348  1.      -0.     
		9.7      0.       1.       0.61371 38.01675  1.      -0.     
		9.8      0.       1.       0.61371 39.66883  1.      -0.     
		9.9      0.       1.       0.61371 41.3927   1.      -0.     
	   10.       0.       1.       0.61371 43.19149  1.      -0.     
	}\TYZPQJK
	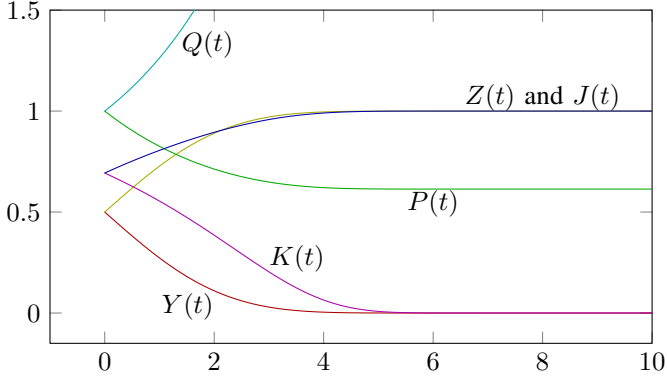
\begin{figure}
		\hskip0cm minus1cm
		\begin{tikzpicture}
			\begin{axis}[height=6cm, width=9.5cm, no marks, xmax=10, ymax=1.5]
				\tikzset{hsb/.code={
					\pgfmathsetmacro\hue{#1}
					\definecolor{this}{hsb}{\hue,1,0.666}
					\pgfkeysalso{this}
				}}
				\addplot [hsb=0/6] table [x=t, y=Y] {\TYZPQJK};
				\addplot [hsb=1/6] table [x=t, y=Z] {\TYZPQJK};
				\addplot [hsb=2/6] table [x=t, y=P] {\TYZPQJK};
				\addplot [hsb=3/6] table [x=t, y=Q] {\TYZPQJK};
				\addplot [hsb=4/6] table [x=t, y=J] {\TYZPQJK};
				\addplot [hsb=5/6] table [x=t, y=K] {\TYZPQJK};
				\path [inner sep=0]
					(1.4,1.4)node[below right]{$Q(t)$}
					(8,1)node[above]{$Z(t)$ and $J(t)$}
					(6,0.61)node[below]{$P(t)$}
					(3,0.2)node[above right]{$K(t)$}
					(2,0.1)node[below left]{$Y(t)$};
			\end{axis}
		\end{tikzpicture}
		\caption{The numerical solution of IVP~\ref{ivp:six}.  Horizontal axis
		is $t$.}                                                 \label{fig:ivp}
	\end{figure}

	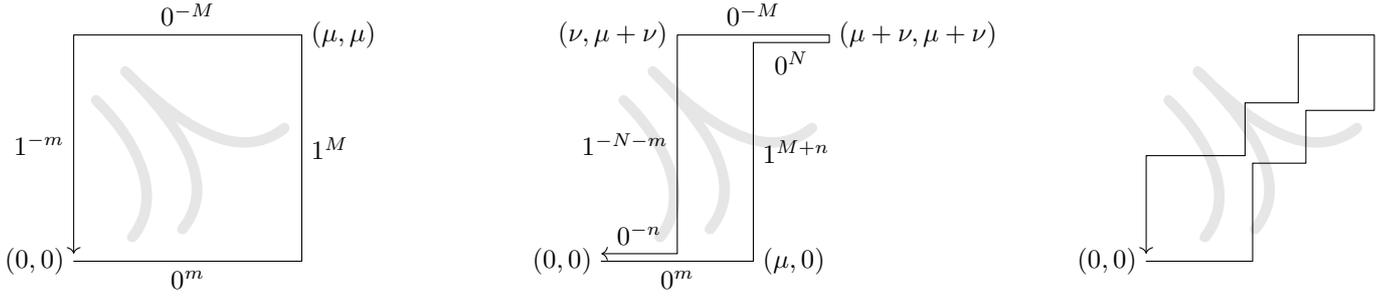
\begin{figure*}
		\centering
		\begin{tikzpicture}[baseline=0]
			\draw[->] (0,0)node[left]{$(0, 0)$}
				-- node[below]{$0^m$} (3,0)
				-- node[right]{$1^M$}
				(3,3)node[right]{$(\mu, \mu)$}
				-- node[above]{$0^{-M}$} (0,3)
				-- node[left]{$1^{-m}$} (0,.1);
			\path[overlay] (1.5,1.5)
				node[rotate=-45,opacity=.1,scale=10]{$\<\,$};
		\end{tikzpicture}
		\hfill
		\begin{tikzpicture}[baseline=0]
			\draw[->] (0,0)node[left]{$(0, 0)$}
				-- node[below]{$0^m$}
				(2,0)node[right]{$(\mu, 0)$}
				-- node[right]{$1^{M+n}$} (2,2.9)
				-- node[below]{$0^N$}
				(3,2.9) -- (3,3)node[right]{$(\mu+\nu, \mu+\nu)$}
				-- node[above]{$0^{-M}$}
				(1,3)node[left]{$(\nu, \mu+\nu)$}
				-- node[left]{$1^{-N-m}$} (1,0.1)
				-- node[above]{$0^{-n}$} (0,.1);
			\path[overlay] (1.5,1.5)
				node[rotate=-45,opacity=.1,scale=10]{$\<\,$};
		\end{tikzpicture}
		\hfill
		\hskip-1cm
		\begin{tikzpicture}[baseline=0]
			\draw[->] (0,0)node[left]{$(0, 0)$}
				-| (1.4,1.3) -| (2.1,2.0) -| (3,3)
				-| (2.0,2.1) -| (1.3,1.4) -| (0,.1);
			\path[overlay] (1.5,1.5)
				node[rotate=-45,opacity=.1,scale=10]{$\<\,$};
		\end{tikzpicture}
		\caption{Three loops that yield interesting results.  Here, $m =
		\int_0^\mu P(t) \,dt = \int_{-\mu}^0 Q(t) \,dt$ and $M = \int_0^\mu Q(t)
		\,dt = \int_{-\mu}^0 P(t) \,dt$.  Same for $n$, $N$, and $\nu$.
		}                                                      \label{fig:paths}
	\end{figure*}

	Local condition determined, we now want to construct a large, fine grid that
	looks like Figure~\ref{fig:10>01}, except that every small square represents
	a short inequality of the form $0^p 1^q \> 1^r 0^s$.
	
	Figure~\ref{fig:01>10} is a numerical example we found on a computer.  In
	this figure, every small square, for instance the lower left one,
	corresponds to a local inequality of the form $0^{1.715} 1^{1.259} \>
	1^{1.015} 0^{2.566}$.  The ideal grid must satisfy two conditions.  (A) The
	gap size must be infinitesimally small and the local condition must hold.
	That is, the delta as defined in the last section should be nonnegative but
	is zero at some point.  (B) The value $y$ such that $\Delta(y) = 0$ must
	aligns.  Let us elaborate on (B).

	Suppose there is an inequality $f(x) \geq g(x)$ and the equality holds for
	$x = 1/3$.  Suppose there is $g(x) \geq h(x)$ and the equality holds for $x
	= 2/3$.  Then $f(x) > g(x)$, but the equality never holds.  In order to
	obtain the tightest inequality, i.e., the equality must hold at some point,
	it is inevitable that we aligns the points at which the sub-inequalities
	assume equality.
	
	Since the point of equality changes continuously, it is best described by
	ordinary differential equations (ODEs).  To solve ODE, we invoke the
	Picard--Lindelöf theorem.

	\begin{lemma}[Picard--Lindelöf]                               \label{lem:PL}
		Let $n \in \mathbb N$.  Let $I \subset \mathbb R$ be an open interval.
		Let $U \subset \mathbb R^n$ be an open rectangle.  Let $(t_0, y_0)$ be a
		point in $I \times U$.  Suppose that a function $F\colon I \times U \to
		\mathbb R^n$ is continuous in the first argument and Lipschitz
		continuous in the second argument.  Then the initial value problem
		\begin{align*}
			y(t_0)           & = y_0,                                          &
			\frac d{dt} y(t) & = F(t, y(t))
		\end{align*}
		has a unique solution for $t - t_0 \in [-\varepsilon, \varepsilon]$ for
		some very small $\varepsilon > 0$.  (Proofs can be found on the
		internet.)
	\end{lemma}

	Using Lemma~\ref{lem:PL} one can prove that the following \emph{initial
	value problem} (IVP) has a unique solution for sufficiently small $t$.
	Using some continuation techniques one can prove that the following IVP has
	na solution for all $t \in \mathbb R$.

	\begin{ivp}                                                  \label{ivp:six}
		Let $(Y, Z, P, Q, J, K)$ be real-valued functions in $t \in \mathbb R$
		satisfying
		\def\ddt{\frac d{dt}}
		\begin{align*}
			Y(0) & = \frac12, & \ddt Y & =   \ln(2) P Y \ln Y, \\
			Z(0) & = \frac12, & \ddt Z & = - \ln(2) Q Z \ln Z, \\
			P(0) & = 1,       & \ddt P & = - \ln(2) (1-J) P Q, \\
			Q(0) & = 1,       & \ddt Q & =   \ln(2) (1-K) P Q, \\
			J(0) & = \ln2,    &     -1 & = \frac Y{\ln Z} \cdot J    
			                             + \frac Z{\ln Y} \cdot K, \\
			K(0) & = \ln2,    &      0 & = \frac{Z\ln Z+Y}{Z\ln(Z)^2} \cdot J 
			                             - \frac{Y\ln Y+Z}{Y\ln(Y)^2} \cdot K.
		\end{align*}
	\end{ivp}
	
	Translation: $t$ points to the southeast direction of the grid.  Functions
	$P$ and $Q$ are the location-dependent version of $p$ and $q$ as in $0^p$
	and $1^q$.  Functions $J$ and $K$ are the location-dependent version of $j$
	and $k$ as in $s - p = \ln(2) \* (1 - j) pq$ and $q - r = \ln(2) \* (1 - k)
	pq$.  Function $Y$ will keep track of the $y$ value in the context of
	equation~\eqref{equ:linear}.  Function $Z$ will be $1 - Y$.  The last two
	equations are just \eqref{equ:linear}.

	\begin{proposition}                                     \label{pro:complete}
		In IVP~\ref{ivp:six}, $Y(t) + Z(t) = 1$ whenever defined.
	\end{proposition}

	\begin{IEEEproof}
		$Y(t) + Z(t)$ satisfies a second-order ODE with initial values $Y(0) +
		Z(0) = 1$ and $Y'(0) + Z'(0) = 0$, thereby being constantly zero.
	\end{IEEEproof}

	\begin{theorem}                                         \label{thm:continue}
		IVP~\ref{ivp:six} admits a unique solution for all $t \in \mathbb R$.
	\end{theorem}

	\begin{IEEEproof}
		Thanks to symmetry it suffices to consider $t \geq 0$.  Throughout the
		entire course of the dynamic system, the following bounds are maintained
		\begin{align*}
			Y(t) & \in \bigl[ 2^{-2^t}, 1 - 2^{-2^{-t}} \bigr],   & 
			Z(t) & \in \bigl[ 2^{-2^{-t}}, 1 - 2^{-2^t} \bigr],   \\
			P(t) & \in \bigl[ e^{1-2^t}, 1 \bigr],                & 
			Q(t) & \in \bigl[       1, 2^t \bigr],                \\
			J(t) & \in \bigl[ k(2^{-2^{-t}}), j(2^{-2^t}) \bigr], & 
			K(t) & \in \bigl[ k(2^{-2^t}), j(2^{-2^{-t}}) \bigr].   
		\end{align*}
		Hence Picard--Lindelöf always applies.  Every time Picard--Lindelöf is
		applied, the \emph{domain of definition} (DoD) extends by a small but
		positive amount.  By Zorn's lemma, the greatest DoD exists.  If the
		greatest DoD is bounded, apply Picard--Lindelöf again to extend that DoD
		and get a contradiction.  The greatest DoD, thereby, must cover the real
		line in its entirety.
	\end{IEEEproof}

	Forgive us for pausing here and appreciating how much trouble the preceding
	theorem has bypassed.  When we applied the standard Runge--Kutta method to
	IVP~\ref{ivp:six}, the solution explodes at about $t = 7.5$.  This, as we
	perceive it, is because $Q(t)$ is about $1.4^t$ and so $Y(t)$ decays to zero
	doubly exponentially fast, which causes the machine to struggle with $\ln
	Z(t)$.

	\begin{definition}[path is the new string]
		Suppose $\Gamma(s) = (u(s), v(s)) \in \mathbb R^2$ is a piecewise smooth
		path parametrized by $s \in [0, S]$.  Treat $t$ as a function in $s$ via
		$t(s) \coloneqq u(s) - v(s)$.  Treat $P$ and $Q$ as functions in $s$ via
		$P(t(s))$ and $Q(t(s))$.  Consider this initial value problem of $X(s)$:
		\begin{align*}
			X(0)          & = x \in [0,1], \\
			\frac{dX}{ds} & = P X\ln(X) \cdot \frac{du}{ds}
			                - (1-X)Q\ln(1-X) \cdot \frac{dv}{ds}.
		\end{align*}
		Define $I_\Gamma(x)$ to be $X(S)$.
	\end{definition}

	In plain text, the function $X$ is defined as follows.  Suppose $\Gamma$ is
	a path on the $u$--$v$ plane.  Suppose at any point $\Gamma(s) \in \mathbb
	R^2$ on the path there associates a value $X(s) \in (0, 1)$.  The values are
	such that, every time $\Gamma$ goes rightward by a tiny amount $du$, the
	associated value $X(s)$ evolves as $X(s+ds) = I_0^{du} (X(s))$.  And every
	time $\Gamma$ goes upward by a tiny amount $dv$, the associated value $X(s)$
	evolves as $X(s+ds) = I_1^{dv}(X(s))$.  Let $I_\Gamma$ be the function
	composition of the infinitesimal $I$'s.

	\begin{proposition}[travel with $Y$]                    \label{pro:together}
		Let $\Gamma(s) = (u(s), v(s)) \in \mathbb R^2$ be a piecewise smooth
		path parametrized by $s \in [0,S]$.  Let $t(s) \coloneqq u(s) - v(s)$.
		Then it holds that $I_\Gamma(Y(t(0))) = Y(t(S))$.  In particular, if
		$\Gamma$ is such that $t(0) = t(S) = 0$, then $I_\Gamma(1/2) = 1/2$.
	\end{proposition}

	\begin{IEEEproof}
		$X(s)$ and $Y(t(s))$ satisfy the same ODE and share the same initial
		condition.
	\end{IEEEproof}
 
	\begin{theorem}[``Green's theorem'']                       \label{thm:green}
		$\Gamma(s) = (u(s), v(s)) \in \mathbb R^2$ is a piecewise smooth path
		that returns back to where it starts, i.e., $\Gamma(0) = \Gamma(S)$, but
		does not self-intersect elsewhere.  Then $I_\Gamma(x) \geq x$
		(respectively, $I_\Gamma(x) \leq x$) if $\Gamma$ goes counterclockwise
		(respectively, clockwise).  Moreover, the equality holds if $x =
		Y(u(0)-v(0))$.
	\end{theorem}

	\begin{IEEEproof}
		Imagine a grid that looks like Figure~\ref{fig:01>10} but the gap size
		$\delta$ is so small that we can safely ignore the cubic error term.
		Put $0^{P(u-v) \delta}$ on the vertical edge centered at $(u, v)$ and
		$1^{Q(u-v) \delta}$ on the horizontal edge centered at $(u, v)$.  Then,
		for any small square with south edge $0^p$, east edge $1^q$, west edge
		$1^r$, north edge $0^s$, we know $I_0^p (I_1^q (I_0^{-s} (I_1^{-r} (x) )
		) ) \geq 0$ (up to cubic error), and the equality holds if $x$ equals
		the value of $Y$ at that point.  Now $I_\Gamma(x)$ is just the grand
		total of all $I_0^p (I_1^q (I_0^{-s} (I_1^{-r} (x) ) ) )$.  So
		$I_\Gamma(x) \geq 0$ and the equality holds if $x$ equals the value of
		$Y$ at the starting point.
	\end{IEEEproof}

\section{Proof of the Main Theorem (Theorem~\ref{thm:main})
         and Three More Consequences}                           \label{sec:main}

	The main theorem states that, if $m$ and $n$ are positive real numbers,
	$0^m 1^n \> 1^m 0^n$ if and only if $(1 - 1/2^{2^m}) ^{2^n} \leq 1/2$.
	Below is the proof.

	\begin{IEEEproof}
		Consider the path on the left of Figure~\ref{fig:paths}.  Applying
		Green's theorem's generalization (Theorem~\ref{thm:green}) to this path
		yields inequalities of the form $0^m 1^M 0^{-M} 1^{-m} \>$ empty string.
		Here, $m = \int_0^\mu P(t) \,dt$ and $M = \int_0^\mu Q(t) \,dt$.  By
		Proposition~\ref{pro:together}, $(1 - 1/2^{2^m}) ^{2^M} = 1/2$.  Thus,
		$(1 - 1/2^{2^m}) ^{2^n} \leq 1/2$ iff $n \geq M$ iff $0^m 1^n 0^{-n}
		1^{-m} \>$ empty string iff $0^m 1^n \> 1^m 0^n$.  This finishes the
		proof of the main theorem, Theorem~\ref{thm:main}.
	\end{IEEEproof}

	In the remainder of this section, let us list three more consequences.

	\begin{corollary}[square]                                 \label{cor:square}
		$0^m 1^M \> 1^m 0^M$ for any positive real number $m$ and $M \coloneqq
		2^m + \log_2(\ln2)$.  Note that $\log_2(\ln2) \approx -0.528$.
	\end{corollary}

	\begin{IEEEproof}
		$(1 - 1/2^{2^m}) ^{2^M} \leq \exp(- 2^M / 2^{2^m}) \leq 1/2$
	\end{IEEEproof}
	
	\begin{corollary}[lightning]                           \label{cor:lightning}
		Suppose $m > n > 0$ are real numbers.  Let $M$ be a function in $m$
		determined by the equation $(1 - 1/2^{2^m}) ^{2^M} = 1/2$.  Let $N$ be a
		function in $n$ determined by the equation $(1 - 1/2^{2^n}) ^{2^N} =
		1/2$.  Then $0^m 1^M 1^n 0^N \> 0^n 1^N 1^m 0^M$, which implies $0^{m-n}
		1^{M+n} \> 1^{N+m} 0^{M-N}$.
	\end{corollary}
	
	\begin{IEEEproof}
		Use the path at the center of Figure~\ref{fig:paths}.
	\end{IEEEproof}

	\begin{corollary}[Dyck Path]                                \label{cor:dyck}
		Let $n > 1$ be an integer.  Suppose $a_j \in \{0, 1\}$ and $b_j = 1 -
		a_j$ for all $j \in [n]$.  Then $a_1 \dotsm a_n \> b_1 \dotsm b_n$ if
		$I_{a_1 \dotsm a_m} (1/2) \leq 1/2$ for any  $m \in [n-1]$ and $I_{a_1
		\dotsm a_n} (1/2) \geq 1/2$.
	\end{corollary}

	\begin{IEEEproof}
		Use a diagonally-symmetric path that looks like the right one in
		Figure~\ref{fig:paths}.
	\end{IEEEproof}

	Mentioned in the introduction, $011 \> 100$ and $001111 \> 110000$ are
	consequences of Corollary~\ref{cor:square}; $00111 \> 10000$ and $000111 \>
	100000$ are consequences of Corollary~\ref{cor:lightning}; and $01011 \>
	10100$ is a consequence of Corollary~\ref{cor:dyck}.

\section{Conclusions}

	We confirms a conjecture that concerns how to compare bit channels in the
	construction of polar code.  The technique is quite novel as we use the
	existence of solution, not the solution per se, of an IVP.  Our results give
	quantitative characterizations of how far away polar coder is from
	Reed--Muller code \cite{Ari08, Ari10, MHU14, LST14, AbY20}.

\bibliographystyle{IEEEtran}
\bibliography{Polder-ODE-25.bib}

\appendices

\section{Proof of Lemma~\ref{lem:g'' and h''}}            \label{pf:g'' and h''}

	Lemma~\ref{lem:g'' and h''} states that $g''(y) > 0$ and $h''(y) > 0$.
	
	\begin{IEEEproof}
		Differentiate $y / \ln(1-y)$ twice, it remains to check whether $-\ln
		(1-y) \geq y / (1-y/2)$.  Compute the Taylor series of both sides at $y
		= 0$ and compare coefficients.  The left-hand side is $1/2, 1/3, 1/4,
		1/5$, etc.  The right-hand side is $1/2, 1/4, 1/8, 1/16$, etc.  That
		will conclude the convexity of $g$.  For the convexity of $h$, invoke
		$h(y) = g(1-y)$.
	\end{IEEEproof}

\section{Proof of Lemma~\ref{lem:g' and h'}}                \label{pf:g' and h'}

	Lemma~\ref{lem:g' and h'} states that $g'(y) > 0$ and $h'(y) < 0$.

	\begin{IEEEproof}
		As $y \to 0$ from the right, $g'(y) \to 1/2$ and is ever increasing.
		Hence $g'$ is positive.  For the positivity of $h'$, invoke $h(y) =
		g(1-y)$.
	\end{IEEEproof}

\section{Proof of Lemma~\ref{lem:j and k}}                    \label{pf:j and k}

	Lemma~\ref{lem:j and k} states that $j(y) > 0$ and $k(y) > 0$

	\begin{IEEEproof}
		Look at equation~\eqref{equ:linear}.  In the second row of the matrix,
		the derivatives $g'(y)$ and $h'(y)$ are positive and negative,
		respectively.  Hence $j$ and $k$ assumes the same sign.  Now look at the
		first row; both $g(y)$ and $h(y)$ are negative.  Hence $j$ and $k$ are
		both positive.
	\end{IEEEproof}

\section{Proof of Lemma~\ref{lem:j' and k'}}                \label{pf:j' and k'}

	Lemma~\ref{lem:j' and k'} states that $j'(y) < 0$ and $k'(y) > 0$ and that
	$j(0) = K(1) = 1$ and $j(1) = k(0) = 0$.
	
	and $k(y) > 0$

	\begin{IEEEproof}
		Look at equation~\eqref{equ:linear}.  Differentiating the first equation
		with respect to $y$, we get $gj' + hk' + h'j + h'k= 0$.  Since $g'j +
		h'k = 0$, we infer that $gj' + hk' = 0$.  As an aftereffect of $g$ and
		$h$ assuming the same sign (they are both negative), $j'$ and $k'$
		assume opposite signs.  Differentiating the second equation with respect
		to $y$, we get $g'j' + h'k' + g''j + h''k = 0$.  As we know that $g''j +
		h''k > 0$ and $(g'j')(h'k') > 0$, it must be the case that $g'j'$ and
		$h'k'$ are both negative.  This proves that $j$ is strictly
		monotonically decreasing and $k$ is strictly monotonically increasing.
		To finalize the proof, notice that $(j,k) = (1,0)$ at $y = 0$ and $(j,k)
		= (0,1)$ at $y = 1$.
	\end{IEEEproof}

\section{More Details of Proposition~\ref{pro:complete}}

	It is clear that $(Y+Z) (0) = 1$ and $(Y+Z)' (0) = (\ln(2)PY\ln Y -
	\ln(2)QZ\ln Z) (0) = 0$ by the given initial values.  It remains to find a
	relation between $(Y+Z)''$ and $(Y+Z)'$ and $(Y+Z)$.  To begin, 
	\begin{align*}
		Y''
		& = (\ln(2) PY\ln Y)' \\
		& = \ln(2) P' Y\ln Y + \ln(2) (P\ln Y+P) Y' \\
		& = \ln(2)^2 (J-1)PQ Y\ln Y + \ln(2) (P\ln Y+P) Y', \\
		Z''
		& = -(\ln(2) QZ\ln Z)' \\
		& = -\ln(2) Q' Z\ln Z - \ln(2) (Q\ln Z+Q) Z' \\
		& = \ln(2)^2 (K-1)PQ Z\ln Z - \ln(2) (Q\ln Z+Q) Z'.
	\end{align*}
	We are to add $Y''$ and $Z''$ together.  The sum will contain a sub-formula
	$JY\ln Y + KZ\ln Z$, which can be replaced by $-\ln Y\ln Z$ thanks to the
	fifth equation in IVP~\ref{ivp:six}.  Afterward, we will replace $Y'$ by
	$(Y+Z)' + \ln(2) QZ\ln Z$ and $Z'$ by $(Y+Z)' - \ln(2) PY\ln Y$.  Let us see
	what those steps lead us to.
	\begin{align*}
		\kern1em&\kern-1em
		(Y+Z)'' \\
		& = \ln(2)^2 PQ (JY\ln Y + KZ\ln Z - Y\ln Y - Z\ln Z) \\
		&\qquad + \ln(2) (P\ln Y+P) Y' - \ln(2) (Q\ln Z+Q) Z' \\
		& = \ln(2)^2 PQ(- \ln Y\ln Z - Y\ln Y - Z\ln Z) \\
		&\qquad + \ln(2) (P\ln Y+P) Y' - \ln(2) (Q\ln Z+Q) Z' \\
		& = \ln(2)^2 PQ (-\ln Y\ln Z-Y\ln Y-Z\ln Z) \\
		&\qquad + \ln(2) (P\ln Y + P - Q\ln Z - Q) (Y+Z)' \\
		&\qquad + \ln(2)^2 (P\ln Y+P) QZ\ln Z \\
		&\qquad + \ln(2)^2 (Q\ln Z+Q) PY\ln Y \\
		& = \ln(2)^2 PQ (Y+Z-1)\ln Y\ln Z \\
		&\qquad + \ln(2) (P\ln Y+P-Q\ln Z-Q) (Y+Z)'.
	\end{align*}
	So $(Y+Z)$ does satisfy  a second-order ODE.

\section{More Details of Theorem~\ref{thm:continue}}

	With an assumption that $t \geq 0$, the claimed bounds
	\begin{align*}
		Y(t) & \in \bigl[ 2^{-2^t}, 1 - 2^{-2^{-t}} \bigr],   & 
		Z(t) & \in \bigl[ 2^{-2^{-t}}, 1 - 2^{-2^t} \bigr],   \\
		P(t) & \in \bigl[ e^{1-2^t}, 1 \bigr],                & 
		Q(t) & \in \bigl[       1, 2^t \bigr],                \\
		J(t) & \in \bigl[ k(2^{-2^{-t}}), j(2^{-2^t}) \bigr], & 
		K(t) & \in \bigl[ k(2^{-2^t}), j(2^{-2^{-t}}) \bigr]    
	\end{align*}
	is obtained in the following order.

	Lower bound on $Q$: The starting point is $Q(0) = 1$ and the derivative is
	positive, hence $Q(t) \geq 1$.
	
	Upper bound on $P$: The starting point is $P(0) = 1$ and the derivative is
	negative, hence $P(t) \leq 1$.

	Upper bound on $Q$: Note that $Q' \leq \ln(2) Q$.  By Grönwall's inequality,
	$Q(t) \leq 2^t$.

	Lower bound $P$: Note that $P' \geq - \ln(2) PQ \geq - \ln(2) 2^t P$.  By
	Grönwall's inequality, $P(t) \geq e^{1-2^t}$. 

	Lower bound on $Y$: Rewrite $Y' \geq \ln(2) Y\ln Y$ as $\ln(Y)' \geq \ln(2)
	\ln Y$.  By Grönwall's inequality, $\ln Y \geq - \ln(2) 2^t$.  That is to
	say, $Y(t) \geq 2^{-2^t}$.

	Lower bound on $Z$: Rewrite $Z' \geq - \ln(2) Z\ln Z$ as $\ln(Z)' \geq -
	\ln(2) \ln Z$.  By Grönwall's inequality, $\ln Z \geq - \ln(2) 2^{-t}$.
	That is to say, $Z(t) \geq 2^{-2^{-t}}$.

	Upper bound on $Z$: This one is as simple as $Z(t) = 1 - Y(t) \leq 1 -
	2^{-2^t}$.

	Upper bound on $Y$: This one is as simple as $Y(t) = 1 - Z(t) \leq 1 -
	2^{-2^{-t}}$.

	Upper bound on $J$: Invoke $J(t) = j(Y(t)) \leq j(2^{-2^t})$.

	Lower bound on $K$: Invoke $K(t) = k(Y(t)) \geq k(2^{-2^t})$

	Lower bound on $J$: Invoke $J(t) = j(Y(t)) = k(Z(t)) \geq k(2^{-2^{-t}})$.

	Upper bound on $K$: Invoke $K(t) = k(Y(t)) = j(Z(t)) \leq j(2^{-2^{-t}})$.

\end{document}